\documentclass[aps,prb,twocolumn,showpacs,10pt]{revtex4}
\usepackage[utf8]{inputenc}
\usepackage{graphicx,amsmath}
\usepackage{amssymb}
\usepackage{amsthm}

\begin{document}
\title{Unraveling of the fractional topological phase in one-dimensional flatbands with nontrivial topology}
\author{Jan Carl Budich and Eddy Ardonne}

\affiliation{Department of Physics, Stockholm University, SE-106 91 Stockholm, Sweden}
\date{\today}
\begin{abstract}
We consider a topologically non-trivial flat band structure in one spatial dimension in the presence of nearest and next nearest neighbor Hubbard interaction. The non-interacting band structure is characterized by a symmetry protected topologically quantized Berry phase. At certain fractional fillings, a gapped phase with a filling-dependent ground state degeneracy, and fractionally charged quasi-particles emerges. At filling $1/3$, the ground states carry a fractional Berry phase in the momentum basis. These features at first glance suggest a certain analogy to the fractional quantum Hall scenario in two dimensions. We solve the interacting model analytically in the physically relevant limit of a large band gap in the underlying band structure, the analog of a lowest Landau level projection. Our solution affords a simple physical understanding of the properties of the gapped interacting phase. We pinpoint crucial differences to the fractional quantum Hall case by studying the Berry phase and the entanglement entropy associated with the degenerate ground states. In particular, we conclude that the `fractional topological phase in one-dimensional flatbands' is not a one-dimensional analog of the two-dimensional fractional quantum Hall states, but rather a charge density wave with a nontrivial Berry phase. Finally, the symmetry protected nature of the Berry phase of the interacting phase is demonstrated by explicitly constructing a gapped interpolation to a state with a trivial Berry phase.
\end{abstract}
\maketitle

\section{introduction}
Interacting topological phases that can not be understood at the level of non-interacting models have attracted continued interest since the discovery of the fractional quantum Hall (FQH) effect \cite{StormerFQH,LaughlinState,PrangeGirvin}. FQH states can be observed in partially filled Landau levels, i.e., in systems that would be metallic in the absence of correlations. At certain fractional fillings, the huge phase space for interactions provided by the macroscopic degeneracy at the Fermi level then conspires with the non-trivial topology of the Landau level \cite{TKNN1982,Kohmoto1985} to yield a gapped state which is topologically distinct from any non-interacting two-dimensional (2D) insulator. To make this distinction more precise, the notion of topological order has been introduced by Wen \cite{WenTO}. The simplest FQH state can be observed at $\nu=\frac{1}{3}$~filling. From a phenomenological viewpoint its key differences from the integer quantum Hall (IQH) state \cite{Klitzing1980,Laughlin1981,TKNN1982} observed in a completely filled Landau level are the following:(i) The Hall conductance $\sigma_{xy}$~representing the topological invariant of the IQH state assumes a fractional value, more concretely $\sigma_{xy}=\frac{1}{3}\frac{e^2}{h}$. (ii) If periodic boundary conditions are imposed, the system exhibits a threefold degenerate ground state for $\nu=\frac{1}{3}$. (iii) The elementary excitations of the state are fractionally charged ($q=\frac{e}{3}$~for $\nu=\frac{1}{3}$) and obey fractional statistics ($\theta=\frac{\pi}{3}$~for $\nu=\frac{1}{3}$).\\

In Ref. \onlinecite{Guo1DFTP}, a similar scenario as the one outlined above for the FQH effect has been studied numerically in a 1D system: These authors consider a topologically non-trivial flat band similar to the model introduced by Su, Schrieffer, and Heeger (SSH) \cite{SSH,SSHReview} at rational filling $\nu=\frac{1}{3}$~which is subjected to short-ranged interactions. Their numerical data indicates remarkable similarities to the FQH setting. The quantized Berry phase \cite{ZakPol,HatsugaiQuantizedBerry} playing the role of the topological invariant of the non-interacting 1D band-structure seems to assume fractional values. The system exhibits a ground state degeneracy of three when periodic boundary conditions are applied. The elementary excitations of the system carry fractional charge.\\

In this work, we present an exact solution of the model for the one-dimensional fractional topological phase (1DFTP) discussed in Ref. \onlinecite{Guo1DFTP} in the physically relevant limit of a large band-gap where a projection onto the partially filled lower band is justified. This approach is analogous to the widely used projection onto the lowest Landau level in the FQH case. Our solution affords an intuitive physical interpretation of all the mentioned peculiarities of the 1DFTP and allows us to scrutinize the key differences between the 1DFTP and the FQH scenario at an analytical level. Our main conclusion is that the 1DFTP is not a one-dimensional analog of the 2D fractional quantum Hall states, but rather a topologically non-trivial charge density wave. In addition, even the phase diagram resulting from the competition between a nearest neighbor (NN) and next-nearest neighbor (NNN) interaction can be precisely understood at the level of our exact solution.\\

In agreement with the general relation between Berry phase and entanglement entropy \cite{RyuBerryEntropy}, we find that a fractional von Neumann entropy characterizes the reduced density matrix of a translation-invariant ground state of a bipartite 1DFTP.
However, when calculating the Berry phase of the interacting ground states we find a remarkable difference to the FQH case: In Ref. \onlinecite{Niu1985}, it has been shown that the Hall conductance of a gapped system is insensitive to twisted boundary conditions (TBC). More precisely, the Hall conductance can be expressed as a constant Berry curvature defined on the torus of twisting angles. For the $\nu=\frac{1}{3}$~FQH case, Niu et al. \cite{Niu1985} have shown explicitly, that the fractionalized Hall conductance can be viewed as the Chern number \cite{TKNN1982,Kohmoto1985} over the enlarged torus of twisting angles that encompasses all three degenerate ground states. In the 1D system under investigation in this work, the Berry phase represents the charge polarization of the system \cite{ZakPol} and is defined in terms of the Berry connection rather than the curvature which brings about a certain basis dependence. More precisely, we can find a basis of the ground state manifold in which only one state carries the total Berry phase of $\pi$~and the other two states are independent of the boundary conditions. However, in a translation-invariant basis, each ground state can be assigned a Berry phase of $\frac{\pi}{3}$. This observation is closely related to the well known fact that a charge polarization only has a relative meaning depending on the choice of a reference unit cell \cite{RestaReview}. In contrast, the Hall conductance of a 2D system is a directly observable quantity with an unambiguously defined value.\\

Furthermore, we investigate the entanglement spectrum of the 1DFTP employing the so called particle cut \cite{SchoutensParticleCut} and find remarkable differences to the Laughlin $\nu=\frac{1}{3}$~state \cite{LaughlinState} in the thin torus limit \cite{KarlhedeTT}. In the quantum Hall case,
the number of `entanglement levels' (or equivalently, the rank of the reduced density matrix), is
given by the number of ground states of the Hamiltonian for the Laughlin state itself, with a
reduced number of  particles, but at the original number of flux quanta \cite{sterdyniak2011}.
Thus, the particle entanglement spectrum of the ground state probes the quasi-hole excitations.
Using the exact solution, we show that for the 1DFTP the rank of the reduced density matrix is
in fact much smaller than the number of quasi-hole states, and in this case merely probes the
fermionic nature of the particles.

Another important difference to the 2D case is that topological order in 1D is always symmetry protected as long as particle number is conserved, see for instance, Refs. \onlinecite{Kitaev2001,WenLU,WenGappedSpin,Budich2013}.
We explicitly show the symmetry protected nature of the Berry phase of the 1DFTP by constructing a gapped interpolation to a trivial charge density wave without any polarization, involving the breaking of the protecting chiral symmetry. During this interpolation the total Berry phase of the ground states changes adiabatically from $\pi$~to zero. In contrast, an interpolation preserving the protecting symmetry involves a phase transition at which the Berry phase jumps to zero.\\

We note that the experimental study of the predicted phase diagram for the 1DFTP should be feasible in a synthetic system of cold atoms in an optical lattice. As has been demonstrated very recently, such settings even afford an experimental access to Berry phases \cite{BlochZakMeasurement}\\

\section{Model and analytical solution}
\label{sec:model}
\begin{figure}[bh]
\includegraphics[width=0.6\linewidth]{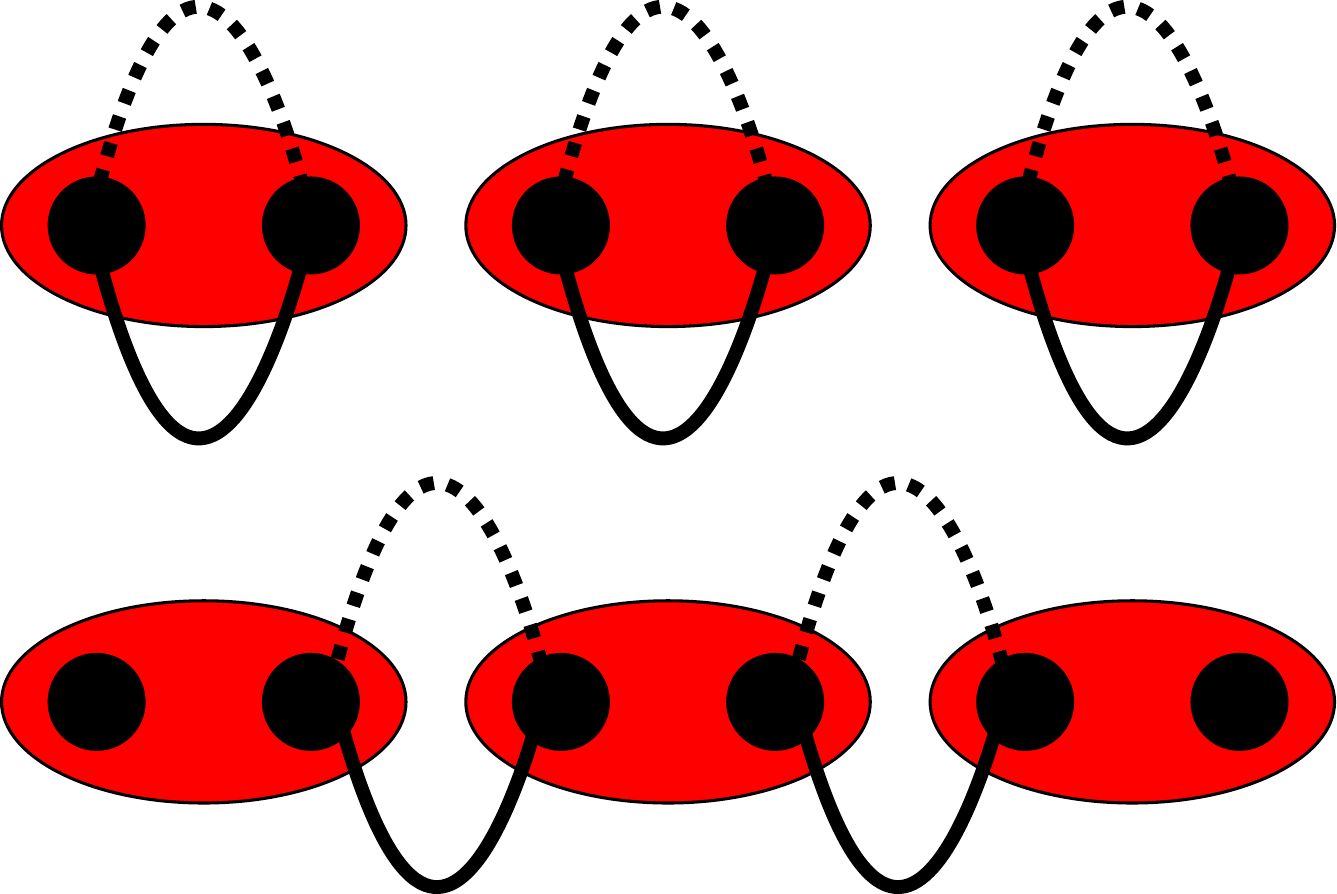}
\caption{Trivial unpolarized model with $v^x=v^z=0\ne v^y$~(top) with orbitals localized on the sites. Non-trivial SSH model with polarized orbitals localized on the bonds between two sites (bottom).}
\label{fig:ssh}
\end{figure}

We consider a two band lattice model similar to the dimer model for polyacetylene originally introduced by SSH in 1979 \cite{SSH,SSHReview}. The tight binding Hamiltonian on which the 1DFTP is constructed reads (see also Ref. \onlinecite{Guo1DFTP})
\begin{align}
H_0=\sum_j c_j^\dag d_{j+1}+d_{j+1}^\dag c_j,
\label{eqn:tightbinding}
\end{align}
where we have chosen unit lattice constant and $c_j,~d_j$~are the annihilation operators of the two orbitals at site $j$. The Bloch Hamiltonian of this model can be written as
\begin{align}
& h_{0}(k)=v^i\sigma_i\nonumber\\
& v^x=\cos(k),~v^y=\sin(k),~v^z=0,
\label{eqn:sshham}
\end{align}
where the $\sigma_i$~denote Pauli matrices in the band pseudo-spin space. The spectrum of this Hamiltonian can be conveniently obtained by taking the square $E_k^2=\lvert v(k)\rvert^2=1$, i.e., the model has completely flat bands. Furthermore, the Hamiltonian anti-commutes with the chiral symmetry operation $\sigma_z$. The chiral symmetry can be viewed as the combination of a particle hole symmetry (PHS) operation $\mathcal C=\sigma_z K$~and the pseudo time reversal symmetry (TRS) operation $\mathcal T=K$, where $K$~denotes complex conjugation. If one of the bands is filled, the system is gapped and its topological invariant $\xi$~is given by the winding number of the map $k\mapsto (v^x(k),v^y(k))^T$~around the origin of the $xy$-plane \cite{Ryu2002}. For our model which is a tight binding analog of the SSH model \cite{SSH,SSHReview}, we obtain $\xi=1$.
The physical consequence of this topologically non-trivial structure is the quantized Berry phase \cite{ZakPol,HatsugaiQuantizedBerry}
\begin{align}
\varphi^B=-i\int_0^{2\pi}\text{d}k~\mathcal A(k)=\pi ~(\text{mod} 2\pi),
\end{align}
where $\mathcal A(k)=\langle l_k\rvert \partial_k \lvert l_k\rangle$~is the Berry connection of the Bloch states $\lvert l_k\rangle$~of the lower band. The Berry phase is related to the charge polarization $P$~ by $2\pi P=\varphi^B$, see Ref.~\onlinecite{ZakPol}. The topological quantization is symmetry protected, i.e., if we allow for $v^z\ne0$~we can adiabatically connect phases with zero and $\pi$~Berry phase. The quantized polarization of the SSH model is illustrated in Fig. \ref{fig:ssh}.
The noninteracting model in Eq. (\ref{eqn:sshham}) is readily diagonalized as
\begin{align}
H_0=\sum_{j,\sigma}\sigma \gamma^\dag_{j,\sigma}\gamma_{j,\sigma},\qquad\sigma=\pm, 
\end{align}
where $\gamma_{j,\pm}=\frac{1}{\sqrt{2}}(c_j\pm d_{j+1})$~are the annihilation operators of the eigenstates.
From now on we consider a fractionally filled lower band and take into account two interaction terms, $V_1$~for NN and $V_2$~for NNN interactions \cite{Guo1DFTP}:
\begin{align}
H_I=V_1\sum_{<i,j>}n_in_j+V_2 \sum_{\ll i,j\gg}n_in_j,
\label{eqn:intham}
\end{align}
where $n_j = c_j^\dag c_j+d_j^\dag d_j$.\\ 

We are interested in the low energy physics coming from the interplay of the macroscopic degeneracy of states at the Fermi energy $E_F=-1$~and the interactions described by $H_I$. Hence, we consider the non-interacting band gap of $2\lvert v\rvert$~as infinitely large compared to the scale of the interaction energies $V_1,V_2$, i.e., a lowest band projection (LBP). This assumption is similar to a projection to the lowest Landau level familiar in FQH physics. From a topological point of view there is a key difference between continuum models (e.g. Landau levels of a homogeneous electron gas in a perpendicular magnetic field) and lattice models. In the Landau level problem, the same Hall conductance of one quantum of conductance can be assigned to all Landau levels. Therefore, the topological defects of the Landau levels only add up to a larger and larger Hall conductance if more and more of them are filled. The physical reason for this is that all eigenstates of the underlying Hamiltonian describe cyclotron motions with the same chirality that is fixed by the direction of the magnetic field. In any lattice model, the situation is fundamentally different. The topology of a lattice model is defined relative to the total Hilbert space spanned by all bands. In more technical terms, the total Hilbert space of all bands can be seen as an embedding space for the bundle of occupied states. Hence, as far as the topological invariant associated with a given energy gap in a lattice Hamiltonian is concerned, the union of all bands above the gap is always the `topological complement' of all states below the gap.      
Along these lines, we argue that all topological features of our lattice model must be encompassed by the LBP approximation. This is because the total Hilbert space of the two bands is topologically trivial so that mixing of the lower band and its `topological complement', i.e., the upper band can only perturb a topological state found in the LBP rather than leading to a richer topological structure.   

%Within the LBP, we rewrite $H_I$~in terms of the $\gamma_{j,\pm}$~operators and discard all terms that contain $\gamma_+$-contributions representing excitations of the upper band.
The LBP amounts to discarding all the $\gamma_+$-contributions to the density operators.
Up to constant energy shifts, the interaction terms then read
\begin{align}
H_I=&\frac{V_1}{4}\sum_j\tilde n_j(2 \tilde n_{j+1}+\tilde n_{j+2})+\nonumber\\
&\frac{V_2}{4}\sum_j\tilde n_j(\tilde n_{j+1}+2 \tilde n_{j+2}+\tilde n_{j+3}) \ ,
\label{eqn:intproj}
\end{align}
where $\tilde{n}_j = \gamma^\dagger_{j,-}\gamma^{\vphantom{\dagger}}_{j,-}$ are the
lower band occupation number operators. 

\begin{figure}[bh]
\includegraphics[width=0.9\linewidth]{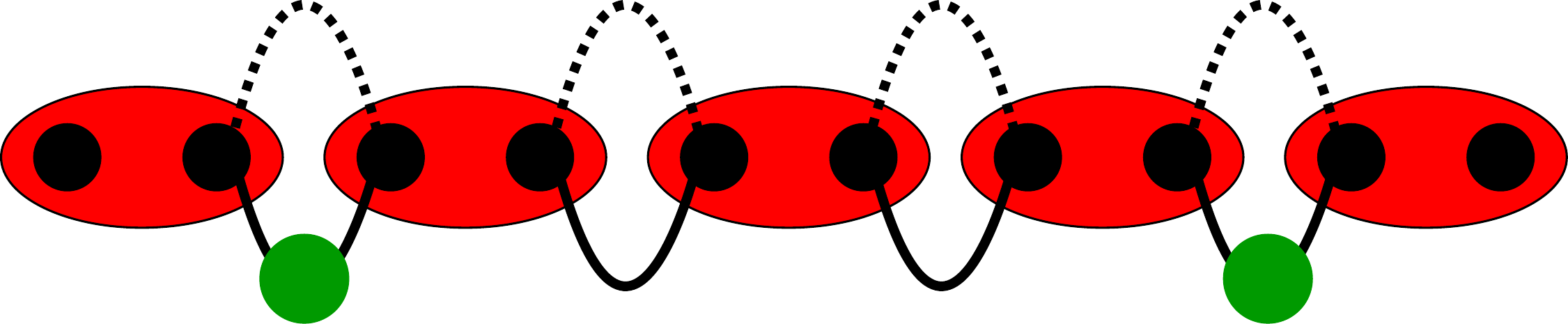}
\caption{Gapped ground state at filling $\nu=\frac{1}{3}$~and $V_1>V_2$. Every third bond is filled with a particle (green dot).}
\label{fig:fracssh}
\end{figure}

At filling $\nu=\frac{1}{3}$~a state with the occupation number pattern $\overline{001}$~obviously annihilates the $V_1$-term which immediately gives the gapped ground state at $V_2=0$~and explains its threefold degeneracy on the ring which has been numerically observed in Ref. \onlinecite{Guo1DFTP}. The degenerate states are obtained by translating the pattern  $\overline{001}$~ (twice) by one lattice site and thus have the patterns $\overline{001},\overline{010},\overline{100}$. The size of the gap is $\frac{V_1}{4}$~as follows immediately from the projected form (\ref{eqn:intproj}) of the interaction Hamiltonian. This state is illustrated in Fig. \ref{fig:fracssh}.
At filling $\nu=\frac{1}{4}$, the state $\overline{0001}$~annihilates both terms in $H_I$~and is ground state with a fourfold degeneracy on a ring and a gap $\frac{V_2}{4}$.\\

A non-trivial situation arises at filling $\nu=\frac{1}{3}$~if both $V_1$~and $V_2$~are non-zero. In this case, placing two defects of the form $0001$~and one of the form $0011$~at constant filling fraction into the ground state $\overline{001}$~might become energetically profitable since the $0001$~string annihilates the $V_2$-term whereas the $001$~pattern does not. Simple counting of interaction energies tells us that this defect pattern saves $\frac{V_2}{2}$~and costs $\frac{V_1}{2}$~due to the $0011$~part. Hence, putting such defects becomes favorable at a critical strength $V_2^c=V_1$~at which the gap closes and a phase transition occurs. We confirmed the appearance of the phase transition at exactly this point in the limit of a large band-gap numerically.

\section{Quantized Berry phase and symmetry protection}
As already mentioned above, the completely filled lowest band of the non-interacting model is characterized by a quantized Berry phase of $\pi$. Employing the method of twisted boundary conditions (TBC) \cite{Niu1985,ProdanTBC}, we would now like to analytically calculate the Berry phase of the gapped interacting phase obtained within the LBP. The notion of TBC can be intuitively understood in the following way. One considers the (arbitrarily large but finite) physical system under investigation as one unit cell of a fictitious super-lattice. The lattice sites of the original lattice are now internal degrees of freedom of the super lattice, i.e., orbitals constituting one super-site. Upon Fourier transforming the super-lattice, each of these orbitals picks up a constant phase $\text{e}^{i\phi X}$, where $X$~labels the super-cell and $\phi$~is the super-lattice momentum. In a particle number conserving Hamiltonian the phase factors of creation and annihilation operators cancel out except for hopping terms crossing the boundary of a super-cell.\\

Let us first apply this program to the non-interacting Hamiltonian (\ref{eqn:tightbinding}) with a super cell of $L$~sites. The model only contains NN hopping. Hence, the only term switching the super cell is the hopping between orbital $L$~of cell $X$~and orbital $1$~of cell $X+1$. The Bloch Hamiltonian of the super-lattice associated with Eq. (\ref{eqn:tightbinding}) is hence given by
\begin{align}
H_{0S}(\phi)= \left(\sum_{j=1}^{L-1}c_j^\dag d_{j+1} + \text{h.c.}\right) + \left(\text{e}^{-i\phi}c_L^\dag d_1 + \text{h.c.}\right).
\label{eqn:supertightbinding}
\end{align}
This model is still readily analytically diagonalized by the operators $\gamma_{j,\pm}(\phi)=\frac{1}{\sqrt{2}}(c_j\pm d_{j+1}),~j=1,\ldots,L-1$~and $\gamma_{L,\pm}(\phi)=\frac{1}{\sqrt{2}}(c_L\pm \text{e}^{-i\phi}d_{1})$. Remarkably, only $\gamma_{L,\pm}$~depends on the momentum variable $\phi$~of the super-lattice. For pedagogical reasons we would like to calculate the Berry phase of this model now in two equivalent ways. First, we interpret Eq. (\ref{eqn:supertightbinding}) as a Bloch Hamiltonian with the occupied bands $\lvert u_{j,-}(\phi)\rangle=\gamma_{j,-}^\dag(\phi)\lvert 0\rangle$. The Berry phase is then readily calculated as
\begin{align}
\varphi^B=&-i\sum_{j=1}^{L}\int_0^{2\pi}\text{d}\phi~\langle u_{j,-}(\phi)\rvert \partial_\phi\lvert u_{j,-}(\phi)\rangle =\nonumber\\
&-i\int_0^{2\pi}\text{d}\phi~\langle 0\rvert \gamma_{L,-}(\phi) \partial_\phi  \gamma_{L,-}^\dag(\phi)\lvert 0\rangle=\pi
\end{align}
An equivalent way to do this calculation is to interpret the Slater determinant $\lvert\Psi_0(\phi)\rangle=\prod_{j}\gamma_{j,-}^\dag(\phi)\lvert 0\rangle$~as the many body ground state of the Hamiltonian (\ref{eqn:supertightbinding}) at half filling. In this many body language the Berry phase can be written as
\begin{align}
 \varphi^B=&-i\int_0^{2\pi}\text{d}\phi~\langle \Psi_0(\phi)\rvert \partial_\phi\lvert \Psi_0(\phi)\rangle=\nonumber\\
&-i\int_0^{2\pi}\text{d}\phi~\langle 0\rvert \gamma_{L,-}(\phi) \partial_\phi  \gamma_{L,-}^\dag(\phi)\lvert 0\rangle=\pi.
\label{eqn:manybodyberry}
\end{align}
The latter approach is more useful for the generalization to the interacting model.\\

\subsection{The interacting model}

We now calculate the Berry phase of the interacting model with the Hamiltonian $H=H_0+H_I$ (see Eq. (\ref{eqn:tightbinding}) and Eq. (\ref{eqn:intham})). As shown in Section \ref{sec:model} the three degenerate exact ground states of the interacting model for $V_2<V_1$~at $1/3$~filling are $\lvert \Psi_l \rangle = \prod_{j=0}^{L/3-1}\gamma_{3j+l,-}^\dag\lvert 0\rangle,~l=1,2,3$. We now again apply TBC. The interaction Hamiltonian $H_I$, in particular in its projected form (see Eq. (\ref{eqn:intproj})), does not depend on the twisting angle $\phi$~since it does not contain any hopping terms. Therefore, even in the presence of TBC, the interacting model can still be solved exactly by just replacing $\gamma_{j,-}$~with $\gamma_{j,-}(\phi)$~in the definition of the ground states $\lvert \Psi_l\rangle$. The resulting ground states $\lvert \Psi_l(\phi)\rangle$~are obviously $2\pi$-periodic in $\phi$~so that the Berry phase defined in total analogy to Eq. (\ref{eqn:manybodyberry}) is a well defined geometric phase. Explicitly, we get
\begin{align}
\varphi_l^B=-i \int_0^{2\pi} \text{d}\phi~ \langle \Psi_l(\phi)\rvert \partial_\phi \lvert \Psi_l(\phi)\rangle=\pi~ \delta_{l,3}.
\end{align}
Two of the ground states thus have zero Berry phase whereas one of them has a Berry phase of $\pi$. This result affords a simple physical interpretation. When imposing the TBC we go to a super-lattice and the Berry phase now describes the polarization of the model in the super cell. Only $\lvert \Psi_3\rangle$~contains an electron which is delocalized over two super-cells, namely the one created by $\gamma_{L,-}^\dag$. Hence $\lvert \Psi_3\rangle$~has a polarization of $\frac{1}{2}$, or equivalently, a Berry phase of $\pi$. In contrast, the two other ground state are unpolarized at the level of the super-lattice description and hence do not contribute to the total Berry phase. If we choose a different basis of ground states, we can distribute the Berry phase differently over the three basis states. For instance, upon combining the ground states $\lvert\Psi_l\rangle$ into momentum states, the latter contribute equally to the Berry phase. However, the total Berry phase $\varphi^B=\sum_l \varphi^B_l$~is always quantized to $\pi$.\\

\subsection{Symmetry protection}

To demonstrate the symmetry protected nature of the present 1DFTP, we now explicitly perform a gapped interpolation between this model and the trivial atomic insulator with the non-interacting Hamiltonian $\tilde H_0=\sum_j c_j^\dag c_j-d_j^\dag d_j$. The Bloch Hamiltonian of this band structure is $\tilde h(k)=\sigma_z$, so it obviously breaks the chiral symmetry. If we subject this model to the same interactions $H_I$~as our original model, we again get three degenerate ground states at $1/3$-filling of the lower band. These states are separated from the rest of the spectrum by a gap $V_2$. We will therefore assume NNN interactions to be present. Since $\tilde H_0$~does not contain any hopping terms this model is completely insensitive to TBC and all three ground states $\lvert \tilde \Psi_l(\phi)\rangle= \prod_{j=0}^{L/3-1}d_{3j+l}^\dag \lvert 0\rangle$~have a zero Berry phase as expected. We now consider the gapped interpolation
\begin{align}
H(\lambda) &= H_0 (\lambda) + H_I \\
H_0 (\lambda) &=\sqrt{1-\lambda}H_0 + \sqrt{\lambda}\tilde H_0 \ . \nonumber
\end{align}
We first note that the Bloch Hamiltonian of the
non-interacting model, $H_0 (\lambda)$, reads
\begin{align}
& h_{\lambda} (k)=v^i\sigma_i \\
& v^x=\sqrt{1-\lambda}\cos(k),~v^y=\sqrt{1-\lambda}\sin(k),~v^z=\sqrt{\lambda} \ .
\nonumber
\label{eqn:intham1}
\end{align}
Thus, this model also has flat bands, because $E_k^2 = |v(k)|^2 = 1$, as was the case
for $H_0$. For both $\lambda = 0$ and $\lambda = 1$, the Hamiltonian has a three-fold degenerate ground state, separated from the excited states by a gap, because we assumed $V_2>0$. The interaction $H_I$ does not depend on the interpolation parameter $\lambda$, and we find that $H(\lambda)$ has a threefold degenerate ground state and a gap for all $0 \leq\lambda\leq 1$.

\begin{figure}[t]
\includegraphics[width=0.8\columnwidth]{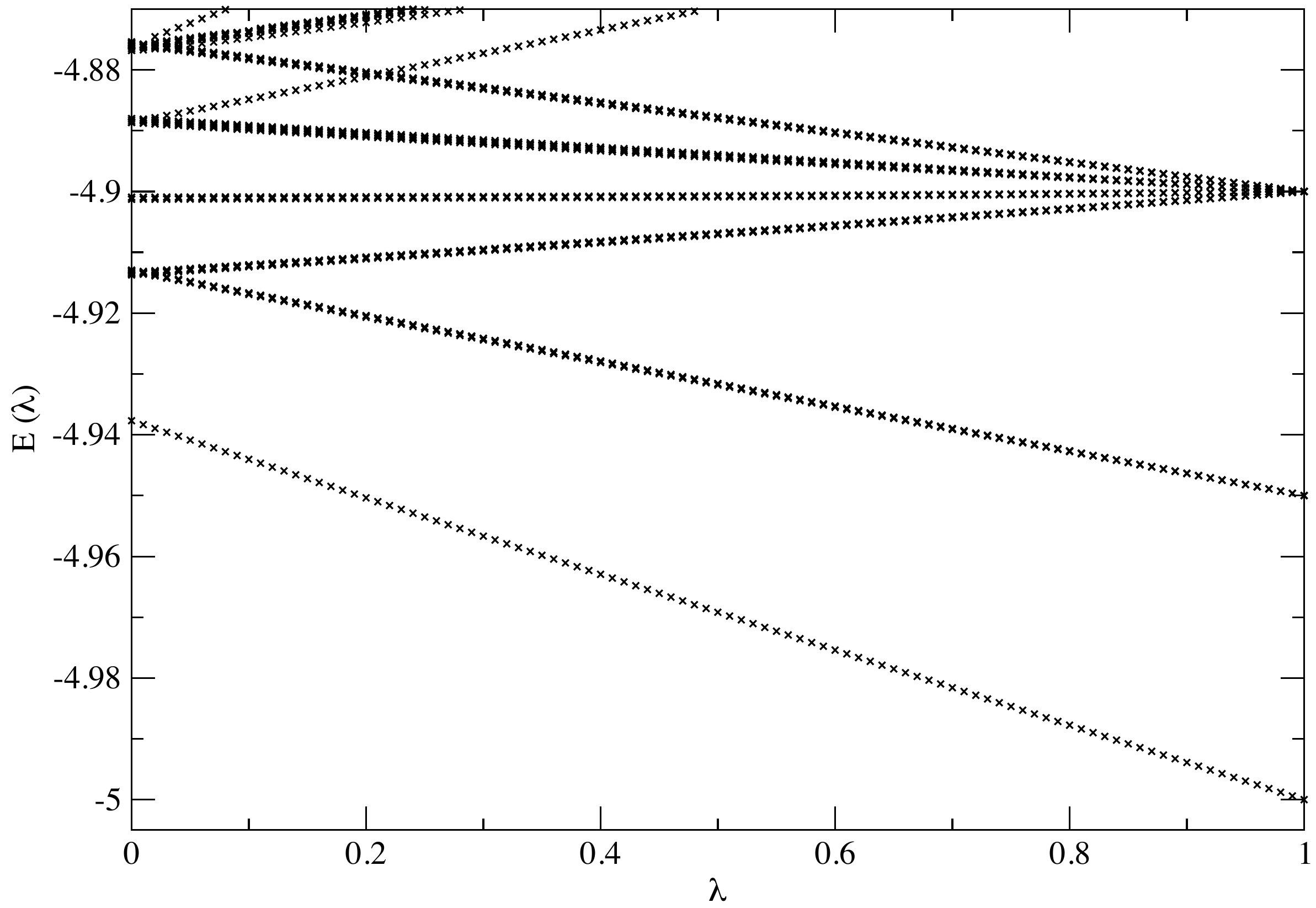}
\caption{The low-lying part of the spectrum of the interacting Hamiltonian $H(\lambda)$, with
interaction parameters $V_1=1/10$, $V_2=1/20$ for system size $L=15$ and periodic
boundary conditions.
We note that the ground state is threefold degenerate for each value of $\lambda$.
}
\label{fig:intspec1}
\end{figure}
In Fig.~\ref{fig:intspec1}, we show the low-energy part of the spectrum for a
system with $L=15$ sites at filling $\nu=1/3$, with
interaction parameters $V_1 = 1/10$ and $V_2 = 1/20$.
We chose these parameters
to be small in comparison to the non-interacting part of the Hamiltonian, so that the
LBP is a good approximation. Indeed, the gap for $\lambda = 0$, namely
$\Delta E \approx 0.02397$ is close to the value $V_1/4 = 0.025$ valid in the LBP
(see section~\ref{sec:model}).
For comparison, the gap for interaction parameters $V_1 = 1/10$ and $V_2 = 0$ is
$\Delta E \approx 0.02423$. The gap for $\lambda = 1$ is $\Delta E = 0.05 = V_2$, the
expected value. We note that our numerical studies are performed on system sizes
similar to the ones used in Ref.~\onlinecite{Guo1DFTP}.

The plot confirms the existence of a gap throughout the interpolation. We only show the case
of non-twisted boundary conditions $\phi = 0$, but the spectrum is in fact gapped for all $\phi$
throughout the interpolation. This is true, because twisting the boundary
conditions only leads to a shift in the momentum $k$ for Hamiltonians conserving
the number of fermions, and the one-particle energies do not depend on $k$. 

\begin{figure}[t]
\includegraphics[width=0.8\columnwidth]{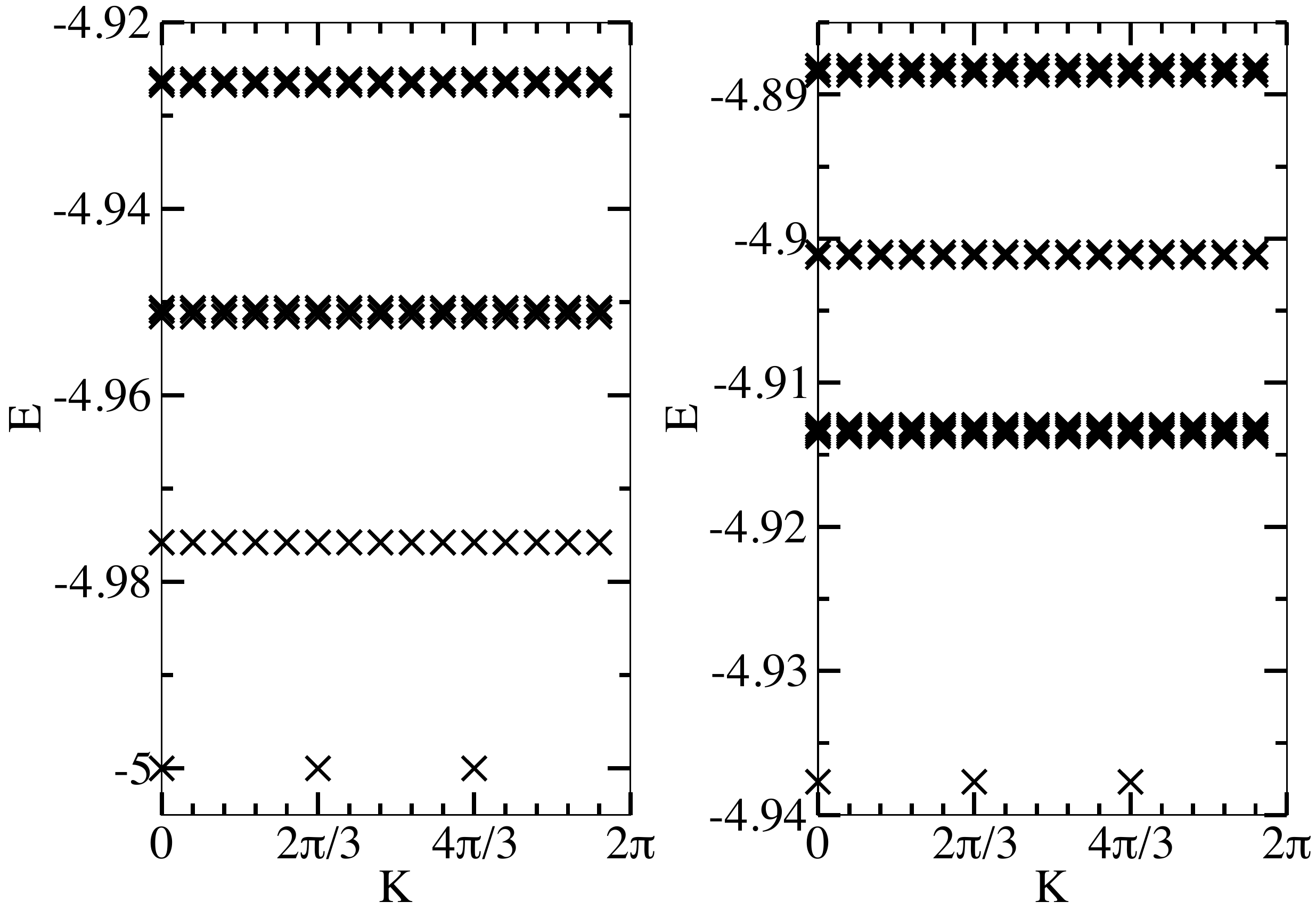}
\caption{The low-lying part of the energy spectrum of the 1DFTP with $L=15$ sites and $F=5$
fermions. The interaction parameters are $V_1=1/10$ and $V_2=0$ (left panel), and
$V_1=1/10$ and $V_2=1/20$ (right panel).}
\label{fig:v2-plots}
\end{figure}

Finally, we note that we checked explicitly that upon increasing $V_2$ from $V_2=0$
to $V_2=1/20$ (while keeping $V_1=1/10$ and $\lambda=0$ fixed), one does not close
the gap. The momentum resolved spectra for the two cases $V_2=0,1/20$ are displayed
in Fig.~\ref{fig:v2-plots}.

\begin{figure}[b]
\includegraphics[width=0.8\columnwidth]{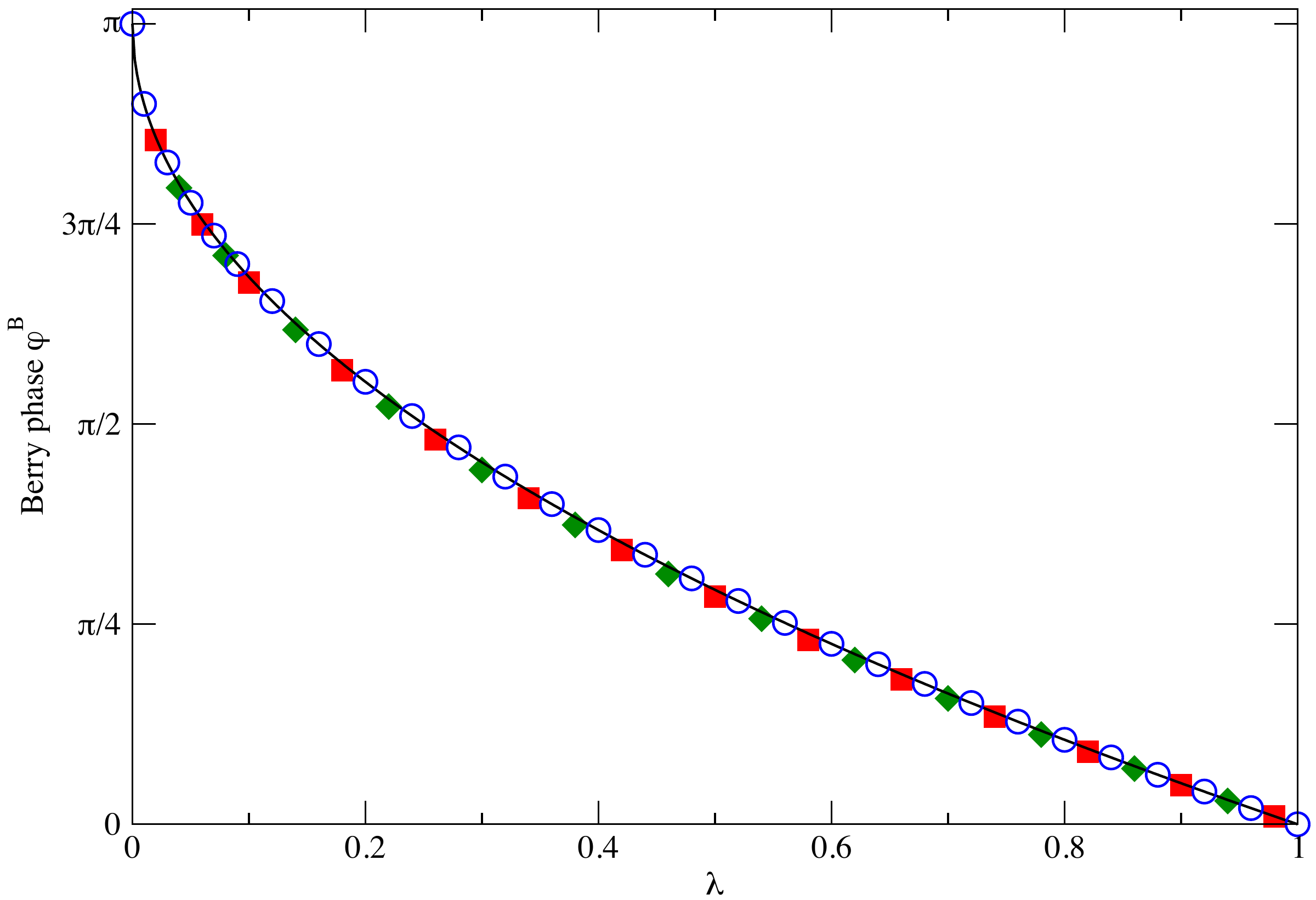}
\caption{%
The numerically calculated Berry phase for the interacting Hamiltonian $H(\lambda)$
with $V_1 = 1/10$,  $V_2 = 1/20$ as a function
of the interpolation parameter $\lambda$ for
$L=3$ (blue circles), $L=6$ (red squares) and $L=9$ (green diamonds).
The black line is the analytic result for the non-interacting case, valid for $L=3$.
}
\label{fig:intberry1}
\end{figure}
To calculate the total Berry phase, we changed the boundary conditions $\phi$ from
$\phi=0$ to $\phi=2\pi$, in steps of $\delta \phi = \frac{2\pi}{100}$.
We considered the system sizes $L=3$ (the non-interacting case), $L=6$ and $L=9$, as
shown in Fig.~\ref{fig:intberry1} by the blue circles, red squares and green diamonds respectively.
The interaction parameters are again $V_1 = 1/10$, $V_2 = 1/20$. In the non-interacting case, the
Berry phase can be calculated analytically. Namely, $\varphi^B$ is given by half the solid angle
mapped out by the curve of $\vec{v}(k)$ on the unit sphere when $k$ changes from
$k=0$ to $k=2\pi$. This leads to $\varphi^{B} = \pi (1-\sqrt{\lambda})$, which is shown as the
black line in Fig.~\ref{fig:intberry1}. We find that the numerically obtained values using exact diagonalization for
$\varphi^B$ at $L=3$ agree perfectly with the analytic result. The Berry phase in the interacting
case for the larger system sizes with $V_1 = 1/10$, $V_2 = 1/20$ deviates only slightly from the non-interacting result, confirming that
the LBP is a good approximation in this regime. 

The Berry phase calculation clearly shows that the quantization of the Berry phase is
protected by the chiral symmetry. Upon breaking this symmetry, we can smoothly change
the Berry phase from $\varphi^B=\pi$ to $\varphi^B=0$, without closing the gap in the spectrum.

\begin{figure}[t]
\includegraphics[width=0.7\columnwidth]{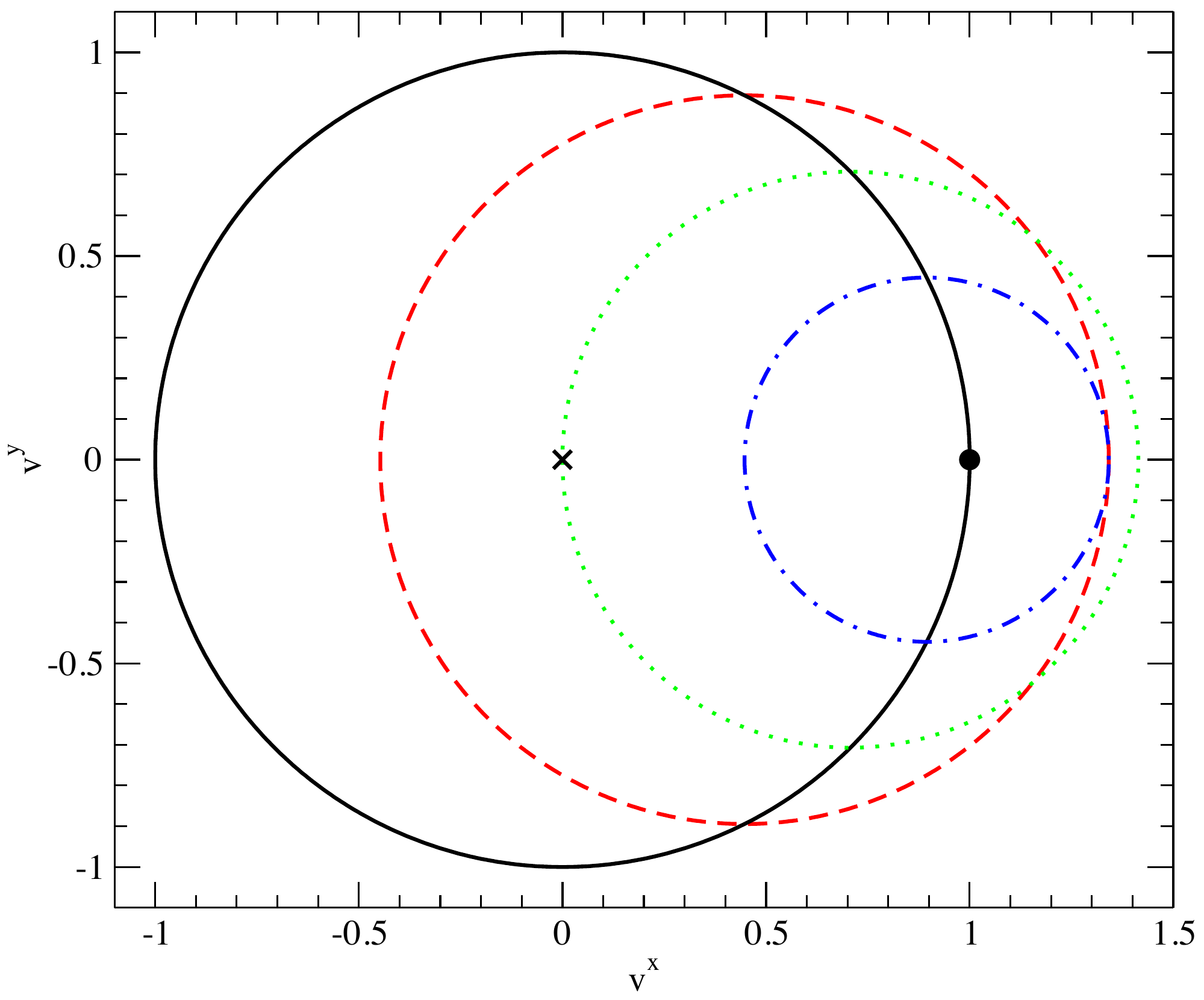}
\caption{%
The map $k\mapsto (v^x(k),v^y(k))$ of Eq.~\eqref{eqn:intham2}, for various values of $\lambda$:
$\lambda = 0$ (black line), 
$\lambda = 1/5$ (red dashed), 
$\lambda = 1/2$ (green dotted), 
$\lambda = 4/5$ (blue dash-dotted), 
$\lambda = 1$ (black dot). The black cross marks the origin, corresponding to
$E=0$, which lies on the curve with $\lambda = 1/2$. 
}
\label{fig:vmap}
\end{figure}

It is also possible to interpolate to a trivial atomic insulator, without breaking the chiral symmetry
(a similar deformation was considered in Ref.~\onlinecite{Guo1DFTP}, in their Fig.~5a).
To this end, we introduce the non-interacting Hamiltonian
$\tilde{H}'_0=\sum_j c_j^\dag d_j + d_j^\dag c_j$, such that the corresponding full interacting
Hamiltonian reads
\begin{align}
H'(\lambda) &= H'_{0} (\lambda) +  H_I \\
H'_{0} (\lambda) &= \sqrt{1-\lambda}H_0 + \sqrt{\lambda}\tilde{H}'_0 \ . \nonumber
\end{align}
The Bloch Hamiltonian of the non-interacting part $H'_{0} (\lambda)$ now reads
\begin{align}
\label{eqn:intham2}
& h'_\lambda (k)=v^i\sigma_i \\
& v^x=\sqrt{1-\lambda}\cos(k)+\sqrt{\lambda},~v^y=\sqrt{1-\lambda}\sin(k),~v^z=0 \ .
\nonumber
\end{align}
Indeed, the chiral symmetry is preserved. The spectrum of the
Bloch Hamiltonian $h'_\lambda (k)$ now depends on the interpolation
parameter $\lambda$ and the momentum $k$, namely
$E_k^2 = 1+2 \sqrt{\lambda(1-\lambda)}\cos(k)$.
In particular, for $\lambda = 1/2$ and $k=\pi$
the gap closes due to a level crossing. We show the map $k\mapsto (v^x(k),v^y(k))$ as
defined by Eq. \eqref{eqn:intham2} for several values of $\lambda$ in
Fig.~\ref{fig:vmap}. The Berry phase $\varphi^B$ corresponding to the lower band is
given by $\pi$ times the winding number of the map $k\mapsto (v^x(k),v^y(k))$ around the origin.
We find that for $\lambda < 1/2$, the Berry phase is $\varphi^B=\pi$, while for $\lambda > 1/2$,
we have $\varphi^B=0$. Precisely for $\lambda = 1/2$, the winding number is not defined,
signaled by the closing of the gap. 

For the interacting Hamiltonian $H'(\lambda)$, the scenario is very much the same, with
the difference that the transition occurs at different values of $\lambda$.
We determined numerically that the transition occurs at
$\lambda \approx 0.34$ for $L=6$ and at
$\lambda \approx 0.27$ for $L=9$. Due to its limited relevance for our main line of reasoning we did not study the precise depence of the gap closing on the system size in more detail here.

\section{Entanglement spectra}

To gain further insight in the nature of the fractional topological phase in one-dimensional
flatbands, we consider both the entanglement entropy and  the entanglement
spectrum. In the study of two-dimensional topological phases, the entanglement
entropy gives insight in the type of topological phase which is realized. Following the work
of Kitaev and Preskill \cite{kitaev2006}, and Levin and Wen \cite{levin2006}, we note that
entanglement entropy $S_{\rm ent}$ associated with dividing the system into two
(real space) regions $A$ and $B$ reads
$S_{\rm ent} = \alpha L - \gamma + O(1/L)$, containing a (non-universal) contribution which
scales as the length $L$ of the boundary, as well as a universal constant
$\gamma = \log \mathcal{D}$.
The quantity $\mathcal{D}$ is the total quantum dimension associated with the topological phase,
which is a measure of the particle content of the topological phase
(see, for instance, Ref.~\onlinecite{kitaev2006a}). In the context of the
fractional quantum Hall states, this universal term has been determined numerically
\cite{SchoutensParticleCut}, and agrees reasonably with the theoretical predictions.

We note that there are several interesting ways
in which one can divide (or cut) the system into two parts $A$ and $B$. The cut considered
in \cite{kitaev2006,levin2006} is the so-called `real space' cut. In the `orbital' cut, one divides the
one-particle orbitals into two sets $A$ and $B$. In the `particle' cut, finally, one divides the
particles into two groups, with $N_A$ and $N_B$ the number of particles in subsystem
$A$ and $B$, respectively. These different ways of dividing the system into two pieces
probes different properties of the system, even though it might not be possible
{\em physically} to actually perform the cut.

The notion of the `entanglement spectrum' was first considered in the context of the quantum
Hall effect by Li and Haldane \cite{li2008}. In short, the entanglement spectrum corresponds to the
full spectrum of the reduced density matrix obtained by tracing out the degrees of freedom in
part $B$. In comparison, the entanglement entropy combines all the eigenvalues into a
single number.

In this paper, we focus on the orbital and particle cuts.
For these cuts, the entanglement spectra for fractional quantum Hall states
were considered for various geometries in previous literature \cite{li2008,lauchli2010,sterdyniak2011}. 

Before we turn our focus on the 1DFTP we study in this paper, we briefly mention
some results concerning the particle entanglement spectrum. We focus on quantum Hall
states for which a model Hamiltonian is known.
This includes the Laughlin states, as well as many non-Abelian
quantum Hall states, such as the Moore-Read state\cite{moore1991}. The reason we focus
on quantum Hall states with a known model Hamiltonian is that one can typically obtain the number
of zero energy ground states of these Hamiltonians, for an arbitrary number of electrons
$N_e$, and an arbitrary number of flux quanta $N_\phi$.
When we divide the electrons into two groups $A$ and $B$, and trace out the electrons
in group $B$, we are left with a system in which the number of particles is reduced,
but the number of flux quanta is unaltered. In the case of model Hamiltonians, one can show
easily \cite{sterdyniak2011} that the rank of the density matrix is bounded by the number of
zero energy ground states of the model Hamiltonian, with $N_A$ electrons but with the original
number of flux quanta $N_\phi$.

It has been observed numerically that for the model quantum Hall states,
this upper bound is indeed reached (see, for instance, Ref.~\onlinecite{sterdyniak2011}).
Proving that the upper bound is reached has turned out to be hard and at presence
a proof is only known for the Laughlin states \cite{garjani2013up}.

In the case of the 1DFTP, we perform a similar analysis by comparing the rank of the
reduced density matrix to the number of ground states for a system with a reduced number
of fermions, but with the same number of sites (playing the role of the number of flux
quanta $N_\phi$).

The rank of the reduced density matrix, or equivalently, the number of levels in the
particle entanglement spectrum, has been used beyond the realm of the fractional quantum
Hall effect. In particular, it has been used to argue for the existence of so-called
two-dimensional `fractional Chern insulators', see \cite{regnault2011} for an early reference.
It was found that the particle entanglement spectrum exhibits a `gap'. The number of states
below this `entanglement gap' was found to be given by the expected value from the
quantum Hall states, showing the relation between the fractional Chern insulators and
the fractional quantum Hall states.

\subsection{The orbital cut}

We briefly discuss the entanglement entropy associated with cutting the system into two
pieces. Because we consider periodic boundary conditions, we effectively cut the system
in two locations when we trace out orbitals in subsystem $B$.

To calculate the entanglement entropy, we start by recalling that if we work in
the LBP, i.e., in terms of the fermions $\gamma_{j,-}$ introduced in Sec.~\ref{sec:model},
the ground states are simple Slater determinants
$\lvert \Psi_l \rangle = \prod_{j=0}^{L/3-1}\gamma_{3j+l,-}^\dag\lvert 0\rangle,~l=1,2,3$. 
Thus, if we perform the cut in terms of the orbitals defined by the fermions $\gamma_{j,-}$,
the entanglement entropy $S_{\rm ent}=0$, because the ground states can be written as
a single product
$\lvert \Psi_l \rangle = \lvert \psi_l \rangle_{A} \otimes \lvert \psi_l \rangle_{B}$.

However, it is more relevant physically to consider the orbitals associated with the original
fermions $c_j$ and $d_j$. In terms of these orbitals, the ground states are not
simple product states, and we will see that the entanglement entropy is nonzero.
We first assume that only NN interactions are present, i.e. $V_2=0$. Then, the
form of the ground states in terms of the fermions $c_j$ and $d_j$ is obtained from
the explicit form of the operators $\gamma_{j,-}$, which are given by
$\gamma_{j,-} = \frac{1}{\sqrt{2}}(c_{j} - d_{j+1})$.
We divide the system into two subsystems $A$ and $B$. If this
division is such that none of the occupied $\gamma_{j,-}$ fermions `has a component'
in both system $A$ and $B$, the entanglement entropy will still be zero, because we can
still write $\lvert \Psi_l \rangle = \lvert \psi_l \rangle_A \otimes \lvert \psi_l \rangle_B$.
If, however, the division is such that one $\gamma_{j,-}$ fermion has a component in both
$A$ and $B$, the entanglement entropy will be given by $S_{\rm ent} = \ln 2$.
Finally, if the cut is such that both boundaries contribute, we find
$S_{\rm ent} = 2\ln 2$ instead \cite{RyuBerryEntropy}.

We now briefly comment on what happens if we increase the interaction parameter $V_2$,
without making the assumption that we are working in the LBP. We will consider the case
that the parameter $V_2$ is small enough, such that we are still in the same phase as for
$V_2 = 0$. In this case, the ground state will have contributions in the upper band,
and is more delocalized in comparison to the case $V_2 = 0$.
This means that upon cutting the system, the ground states will have longer range
correlations across the boundaries, leading to an increased entanglement entropy, and
an increase in the rank of the reduced density matrix.
We confirmed this behavior numerically.

\subsection{The particle cut}

We pointed out above that in the case of quantum Hall states, the rank of the reduced
density matrix, associated with dividing the electrons in two groups, is related to the
number of ground states for the reduced number of particles, but with the same flux.

We therefore now first take a quick look into the number of ground states of the
1DFTP, in case that the filling is reduced from $\nu=1/3$. To deduce the number of ground
states of the 1DFTP, with only nearest-neighbor interactions $V_1$ present, we use exactly the
same arguments as in Sec.~\ref{sec:model}. We consider a system consisting of
$L$ sites, filled with $F$ fermions. We assume that the filling $\nu=\frac{F}{L} \leq \frac{1}{3}$,
such that that there are ground states with the same energy as the
ground states at filling $\nu=\frac{1}{3}$.
The number of ground states is given by the number of ways in which we can distribute the
$F$ fermions over the $L$ sites, such that no three consecutive orbitals contain more than
one fermion. This number of ground states is precisely equal to the number of
quantum Hall states on the torus, in the so-called `thin-torus' limit
\cite{bergholtz2005,bergholtz2006,seidel2006}.
Employing some combinatorics shows that the number of ground states is given explicitly by
$\frac{L}{F} \tbinom{L-2F-1}{F-1}$, if $F>0$ (see also Ref.~\onlinecite{seidel2011}).
For $F=0$, there is trivially only one state. We checked numerically that this indeed gives
the correct number of ground states of the 1DFTP, when only NN interactions are present.

Because we have the exact solution of the 1DFTP model at our disposal, we can determine the
particle entanglement spectrum exactly. As was the case for the orbital cut, it is easiest to
do so in terms of the eigenstates
$\lvert \Psi_l \rangle = \prod_{j=0}^{L/3-1}\gamma_{3j+l,-}^\dag\lvert 0\rangle,~l=1,2,3$.
These states constitute a basis for the threefold degenerate ground state of the model at
filling $\nu = \frac{1}{3}$. Because we divide the system into two parts $A$ and $B$, by
means of dividing the fermions into two groups, we can perform the calculation of the
particle entanglement spectrum directly in the basis of $\gamma_{j,-}$ fermions. Rewriting
these fermions in terms of the original $c_j$ and $d_j$ fermions is, for the present cut,
only a local transformation, which does not change the spectrum of the reduced density
matrix.

We concentrate on the ground state
$\lvert \Psi_3 \rangle = \prod_{j=1}^{L/3}\gamma_{3j,-}^\dag\lvert 0\rangle$, which is the
Slater determinant, such that each third orbital is filled. We now number the fermions
and declare the fermions numbered
$1,\ldots,N_A$ to belong to part $A$, while the remaining fermions belong to part
$B$. Because the ground state $\lvert \Psi_3 \rangle$ is a single Slater determinant,
it follows that the rank of the density matrix is given by the number of ways one can
divide the $N_A$ fermions over the orbitals which are occupied in the original ground
state $\lvert \Psi_3 \rangle = \prod_{j=1}^{L/3}\gamma_{3j,-}^\dag\lvert 0\rangle$ (we refer to
Refs.~\onlinecite{sterdyniak2011,chandran2011} for more details on calculating the particle
entanglement spectrum).

It follows that the number of non-zero eigenvalues of the reduced density matrix is given by
$\tbinom{L/3}{N_A}=\tbinom{F}{N_A}$. Moreover, all these non-zero eigenvalues are equal
to one another. We note that if we had calculated the reduced
density matrix of the momentum eigenstates formed with $\lvert \Psi_l \rangle$, with $l=1,2,3$,
we would have found that the reduced density matrix has $3 \tbinom{F}{N_A}$
degenerate non-zero eigenvalues. 

Having obtained the entanglement spectrum, we can now compare the number of entanglement
levels to the number of ground states of the Hamiltonian at the reduced number of
fermions. Making use of the formula we gave above, we find that the number of
ground states of the Hamiltonian with $L = 3F$ sites, with $N_A>0$ fermions, is given
by $\frac{3F}{N_A} \tbinom{3F-2N_A-1}{N_A-1}$ (or $1$, if $N_A=0$),
which should be compared to the rank of the reduced density matrix $\tbinom{F}{N_A}$.
We find that for $N_A=0$, these numbers are both equal to $1$ trivially. For $N_A>0$,
we find that the rank of the reduced density matrix is lower than the upper bound coming
from the Hamiltonian. In particular, for $N_A=1$, the former is given by $F$, while the
latter is $3F$. For $N_A=F$, the former is $1$, while the latter is $3$. In general, the
ratio of the rank of the reduced density matrix and the upper bound is much smaller
than one-third.

The fact that the rank of the reduced density matrix of the particle entanglement spectrum
is much smaller than the upper bound coming from the Hamiltonian marks a striking
difference with the quantum Hall case, for which this upper bound is in fact satisfied.
Indeed, the `correlations' present in the fractional quantum Hall states, which are
probed by the particle entanglement spectrum, are of a more non-trivial kind. 
In the 1DFTP, they merely signal the fact that the ground states are Slater determinants
of identical fermions.

We close this section by making the following remark. In determining the particle
entanglement spectrum above, we did not make use of the fact that the interacting
fermions are occupying a band with a non-trivial topology. In particular, we can give
exactly the same arguments for the model in which we interpolated the bands to the
trivial atomic insulator. In that case, we obtain exactly the same particle entanglement
spectrum. Thus, the particle entanglement spectrum does not, in the present case, distinguish
between the topological and trivial cases. This is not surprising, because for the
models we study, the particle entanglement spectrum probes the fermionic nature of the particles in
the model.

\section{Concluding discussion}

We investigated the 1DFTP, which was first considered in Ref. \onlinecite{Guo1DFTP}.
We solved the model in the limit where the interactions are small compared to the band gap,
by projecting the model onto the lowest band. This projection is analogous to considering
the quantum Hall effect  `in the lowest Landau level'. The exact solution explains the observed
threefold degenerate ground state and allows for the determination of the phase diagram.
Although the 1DFTP shares certain features of the fractional quantum Hall states, there are
also crucial differences.

The 1DFTP exhibits a threefold degenerate ground state (when considering periodic
boundary conditions), and fractionally charged excitations,
just as is the case for the $\nu=\frac{1}{3}$ fractional quantum Hall state.
We considered the interacting model in a flatband with non-trivial topology (as in
Ref. \onlinecite{Guo1DFTP}), as well as in a trivial flatband. By choosing the appropriate
interpolation (i.e., without breaking the chiral symmetry), we showed that one can
adiabatically interpolate the 1DFTP from the topological to the trivial case.
That way we showed unambiguously that both the ground state degeneracy and the
fractionally charged excitations should not be viewed as emerging because of the
topological band structure, but rather as consequences of the CDW physics
describing both cases. 
We also demonstrated that the Berry phase changes continuously from
$\varphi_B=\pi$ for the topological flatbands to $\varphi_B=0$ for the trivial flatbands, in
the interpolation mentioned above.

Using the exact solution, we analytically calculated the Berry phase associated with the
degenerate ground state in the case of the topological flatbands.
This calculation revealed that upon changing the basis for the threefold
degenerate ground state, one can change the relative contribution from each state to the total
(quantized) Berry phase $\varphi_B = \pi$. This basis dependence can be understood by
realizing that in one-dimensional systems, the Berry phase is a measure of the charge
polarization \cite{ZakPol}, which depends on the choice of the unit cell. Thus, one can not
entertain the objective notion of a fractionalized Berry phase. In contrast, in the two-dimensional
quantum Hall effect, the fractionalized Hall conductance is a physical observable that is directly accesssible experimentally and hence represents objective physical reality.

The basis dependence of the charge polarization alluded to in the previous paragraph also
plays a role in the entanglement entropy associated with dividing the fermionic
orbitals into two sets. We showed that in the basis of $\gamma_{j,-}$ orbitals, appearing in
the exact solution of the model, the ground states are single Slater determinants. Dividing
these orbitals into two sets gives a vanishing entanglement entropy. In terms of the original
orbitals in which the model is phrased, the electrons are delocalized, giving a finite polarization.
Using the CDW basis for the ground state, this in turn gives rise to a finite entanglement entropy
$p\ln 2$, where $p=0,1,2$ is the number of fermions delocalized over the cut.

We also considered the entanglement spectrum associated with dividing the fermions
themselves into two sets. In the context of fractional quantum Hall states (and fractional
Chern insulators), the particle entanglement spectrum of the ground state is directly
related to the excitations of the system. This is signaled by the rank of the reduced
density matrix, which equals the upper bound set by the Hamiltonian of the system
itself. In calculating the entanglement spectrum for the 1DFTP, we found that the rank
of the reduced density matrix is in general much lower, and can be understood
completely by considering the fermionic nature of the particles. This implies that the
ground state of the 1DFTP does not contain the correlations necessary to provide full
knowledge of the excitations of the system, in contrast to the quantum Hall case.

The exact solution revealed that in the presence of NN interactions only, there is a
three-fold degenerate ground state at filling $\nu=\frac{1}{3}$. Upon adding a repulsive
NNN interaction, one finds that there is a (fourfold) degenerate ground state at
filling $\nu=\frac{1}{4}$. Thus, by merely changing the range of the two-body interaction,
we can change the filling at which the ground state occurs from having an odd denominator
to one having an even denominator. In the quantum Hall case, one can not perform such
a simply change in the filling fraction, because the fermionic nature of the electrons does not
allow for a fermionic Laughlin state at even denominator filling.

The Moore-Read quantum Hall state\cite{moore1991} does have even denominator filling
fraction. The model Hamiltonian which has the Moore-Read quantum Hall state as its
ground state is a three-body interaction. In the flatband models we considered in this paper,
one can also consider three-body interactions. In fact, it is straightforward to construct an
electrostatic three-body interaction in terms of the $\gamma_{j,-}$ fermions, which has
the expected six-fold degenerate ground state at filling $\frac{1}{2}$.
Although this model shares some properties with its quantum Hall cousin, it exhibits
the same differences as the 1DFTP in comparison to the Laughlin states.

We note that the entanglement spectra of fractional topological insulators in the one-dimensional `thin-torus' limit has been considered in Ref. \onlinecite{ttfci12up}, which has some overlap with the results presented in this paper.

{\em Acknowledgements}. This research was sponsored, in part, by the swedish research council.
 
%\bibliographystyle{apsrev}
%\bibliography{1dftp}

\end{document}